\documentclass{article}
\usepackage[accepted]{vietnam} 
\usepackage{natbib}
\usepackage{graphicx}      
\newcommand{\sigmac}{{\sigma_c}}

\newcommand{\Mdotc}{{\dot{M}_c}}
\newcommand{\Mc}{{M_c}}
\newcommand{\Rc}{{R_c}}

\newcommand{\Mdotstar}{{\dot{M}_*}}
\newcommand{\Mstar}{{{M}_*}}

\newcommand{\SFRff}{{{\rm SFR}_{\rm ff}}}
\newcommand{\tff}{{t_{\rm ff}}}
\newcommand{\tffc}{{t_{{\rm ff},c}}}

\newcommand{\Sigmac}{{\bar\Sigma_c}}

\newcommand{\vchar}{v_{{\rm ch},w}}

\newcommand{\rch}{r_{\rm ch}}

\newcommand{\ltsim}{\lower.5ex\hbox{$\; \buildrel < \over \sim \;$}}
\newcommand{\gtsim}{\lower.5ex\hbox{$\; \buildrel > \over \sim \;$}}

\begin{document}
\twocolumn[
\title{Accretion and Feedback in Star Cluster Formation}
\titlerunning{Star Clusters: Accretion \& Feedback}
\author{Christopher D. Matzner}{matzner@astro.utoronto.ca}
\address{Dept.\ of Astronomy \& Astrophysics, University of Toronto, Toronto  ON M5S 3H4, Canada}

\vskip 0.5cm 
]

\begin{abstract}
Star cluster formation is unlikely to be a sudden event: instead, matter will flow to a cluster's formation site over an extended period, even as stars form and inject energy to the region.   A cluster's gaseous precursor must persist under the competing influences of accretion and feedback for several crossing times, insofar as star formation is a slow process.   The new-born stellar cluster  should therefore preserve a memory of this competition.  Using analytical approximations we assess the dynamical state of the gas, mapping regimes in which various types of feedback are weak or strong.   Protostellar outflows, radiation pressure, and ionized gas pressure are accounted for.   Comparison to observations shows that feedback is often incapable of expelling gas in the more massive, rapidly-accreting  clusters; but feedback may nevertheless starve accretion by acting on larger scales. 
\end{abstract}

\section{Introduction}

It has long been known that star formation is highly correlated in space and time, with bursts of activity that create a very wide range of stellar associations -- although most of these rapidly disperse.   Star formation is also famously inefficient -- slow (compared to the free fall rate, e.g.\ \citealt{2007ApJ...654..304K}) and ineffectual (compared to the total molecular mass).   At the same time, large reservoirs of dense, non-star-forming gas has been discovered,  often in the form of filamentary structures \citep[e.g.][]{1979ApJS...41...87S,1987ApJ...312L..45B,1989ApJ...338..902L,1999ApJ...510L..49J,2010A&A...518L.102A,2012A&A...540L..11S,2014A&A...561A..83P}.    This, and many observations of cluster formation at the intersections of filaments, suggest a hub-and-spokes arrangement \citep{2009ApJ...700.1609M} in which cluster formation is largely fed by (probably filamentary) accretion.   Rapid accretion onto proto-cluster sites is indeed observed in many cases, including regions where massive stars have already formed \citep[e.g.,][]{2016MNRAS.461.2288H}.

\begin{figure*}
\vskip -0.3cm
\centering
$\begin{array}{cc}
\includegraphics[angle=0,width=13.cm]{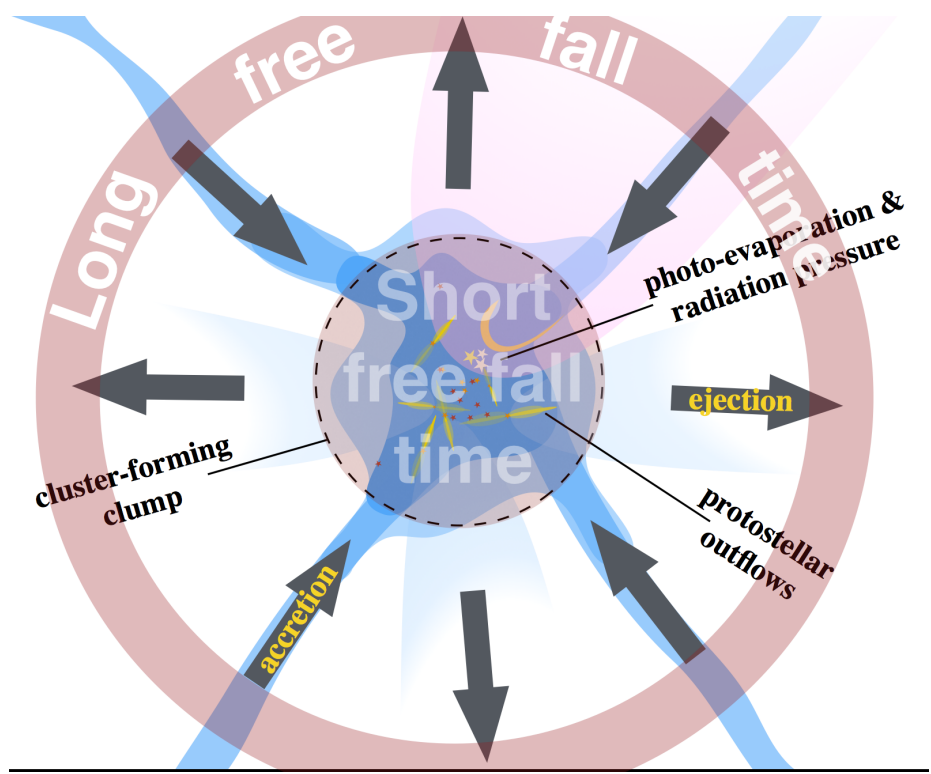} 
\end{array}$
\caption{
Schematic, adapted from MJ15, of an accreting cluster-forming region with energy injection and mass ejection due to stellar feedback.  An extended or filamentary mass reservoir with a relatively long free-fall time accretes onto a central gaseous `clump'.  Because this clump has a shorter free-fall time, but star formation takes many free-fall times, it must persist as a virialized object while being eroded and stirred by protostellar outflows, photo-ionization, stellar winds, radiation pressure, and (eventually) supernovae. }
\label{fig:Schematic}
\vspace{-0.5cm}
\end{figure*}

But the combination of slow star formation, and ongoing accretion from an extended reservoir, poses a puzzle.  A natural duration for the accretion event is the free-fall time $t_{\rm ff} = \sqrt{3\pi/(32 G \bar \rho)}$ of the reservoir.  Whatever this is, it is surely longer than the {\em local} free-fall time of the cluster gas -- first, because the density within a filament is greater than the mean density $\bar\rho$ of the region that contains it; and second, because the cluster-forming region, which accumulates matter from the filaments, tends to be even denser.   Therefore, any gas condensation growing by accretion must persist for multiple internal dynamical times, as argued on other grounds by \citet{2006ApJ...641L.121T}.  See the schematic in Figure \ref{fig:Schematic}.    Consumption of the gas by stars takes even longer: tens of local dynamical times.    What is the state of the gas condensation in the period when matter arrives but cannot escape?  And, how does energetic feedback from stars change this? 

\section{Accretion: a growing clump} 

These questions motivated the analysis of \citeauthor{2015ApJ...815...68M} (\citeyear{2015ApJ...815...68M}; hereafter MJ15), which I summarize here.   Its spirit is to adopt the simplest description that nevertheless accounts for all the key elements (mass accretion, star formation, various forms of feedback).  So, the cluster-forming `clump' is described with a single mass, radius, column, and velocity dispersion ($\Mc, \Rc, \Sigmac=\Mc/(\pi \Rc^2)$, and $\sigmac$), and the infall is described only by the accumulation rate $\Mdotc$.   Many critical details, like the magnetization and clumpiness of the reservoir, are assumed to affect the model parameters discussed below, but are not considered in detail.  
While crude, this strategy suffices to set some baseline expectations for the process.  

For instance, the mass and accretion rate are enough to estimate the clump's radius, velocity dispersion, and column.  The fact that clump gas must linger for multiple dynamical times means that it cannot be far from virial equilibrium: $ \sigmac^2 \simeq G \Mc/(5 \Rc)$.   Moreover, if clump is supported by turbulence (rather than rotation or thermal pressure) and turbulence is generated by inflow (rather than stellar feedback), then the decay of turbulent energy ($\sim \Mc\sigmac^3/\Rc$) must balance the accretion luminosity ($\sim G \Mc \Mdotc/\Rc$) so that $\sigmac^3 \simeq G \Mdotc$.   So, combining these, $\Rc \simeq G^{1/3} \Mc/(5 \Mdotc^{2/3})$.   Steady accretion causes $\Rc$ to rise and $\Sigmac$ to fall at constant $\sigmac$, whereas accelerating accretion leads to an increase in $\sigmac$ (although $\Sigmac$ still falls, unless $\Mdotc$ increases as rapidly as $t^4$).     A shut-off of accretion would break the assumption that accretion luminosity balances turbulent decay; the clump would contract at constant mass in the absence of other influences.  

\section{Star formation and energetic feedback} 

Within this evolving clump environment, stars form at some dimensionless rate $\SFRff = \Mdotstar \tffc/\Mc$  and influence the gas by several means (``feedback'').   The basic assumption of this work is that $\SFRff$ changes only slowly as the clump evolves, and that its value is much less than unity.   While the low average value of $\SFRff$ is not controversial, its relative constancy certainly is.   MJ15 adopt $\SFRff\simeq 1/30$. 

\subsection{Protostellar outflows} 

First among the feedback mechanisms are the protostellar outflows -- jet-driven flows that accompany most or all star formation, injecting a momentum $\Mstar \vchar$ for every new stellar mass $\Mstar$, where $\vchar\sim 10-30\,$km\,s$^{-1}$.     These flows are complicated, being magnetically driven and collimated as well as intermittent.  Although one can account for some of these complications (for instance, in the specific but idealized models by \citealt{2000ApJ...545..364M} and \citealt{2007ApJ...659.1394M}, summarized by MJ15), the simple conclusion is that outflows matter only for low values of $\sigmac$.   The reason is that the outflow-driven acceleration ($\Mdotstar \vchar/\Mc = \SFRff \vchar/\tff \simeq \SFRff \vchar \sigmac/\Rc$), overwhelms the turbulent acceleration ($\sigmac^2/\Rc$) only when $\sigmac \ltsim \SFRff \vchar$, i.e., for $\sigmac$ less than a few km\,s$^{-1}$.   Plenty of embedded clusters have velocity dispersions in this range, including the progenitor of the Orion Nebula Cluster, but the same cannot be said for more massive cluster-forming regions. 

MJ15 incorporate mass loss and energy injection from protostellar outflows into the expectations for $\Sigmac$ and $\sigmac$ in an accreting clump; this is convenient, as these parameters are readily observable and directly related to the influence of massive stars. 

\subsection{Radiation pressure} 

Next, consider feedback due to the massive stars, the simplest of which is the radiation pressure.  This can be broken into two parts, one due to the direct force of starlight ($L_*/c$ for a stellar luminosity $L_*$), and another due to the re-radiation and recapture of the stellar luminosity, primarily by dust grains.   MJ15 take pains to estimate the latter (reprocessing) contribution carefully, but conclude that it only adds a factor of order unity to the direct force.   In this case, radiation pressure is important only at low values of the mean clump column $\Sigmac$.  This stems from the fact that the luminosity per unit mass ($L_*/\Mstar$) is reasonably constant, if mass function-dependent, at about $10^3L_\odot/M_\odot$ for a young stellar population ($t\ltsim 4\,$Myr).  The acceleration of the clump gas by direct radiation is therefore $\sim (L_*/M_*)(M_*/\Mc)c^{-1}$, or $6\times10^{-8}$\,cm\,s$^{-2}$ when stars and gas  share the mass equally.   The clump's surface gravity, on the other hand, is $G\Mc/\Rc^2 = \pi G\Sigmac$, which is lower than the radiation-induced acceleration when $\Sigmac\ltsim 0.3$\,g\,cm$^2$.    Low-column regions should blow themselves apart -- so long as they are old enough, and massive enough, for the massive stars to have formed in the first place.   

MJ15 implement a model in which the clump is patchy (involving a log-normal distribution of columns, as observed in nearby clouds), so that some blowout occurs even at higher $\Sigmac$.  However the slow nature of star formation reduces the typical mass fraction in stars, lowering the critical column to $\sim 0.1$\,g\,cm$^{-2}$. 

\subsection{Photoionization and photo-evaporation} 

Photoionization is the process in which stellar photons ionize neutral gas, heating it to $\sim 10^{3.9}$\,K, and causing either an explosive expansion or an ionized outflow, or both.  Photoionization can severely limit the efficiency of massive star formation on giant molecular cloud scales \citep{1979MNRAS.186...59W}, while also creating copious turbulent energy within giant clouds \citep{2002ApJ...566..302M}.    How important is it during star cluster formation?  

To answer this, MJ15 make use of the fact that dusty photoionized regions can be described to good approximation as a one-parameter family of varying radiation pressure (compared to gas pressure) and varying dust optical depth.   Earlier,  \citet{2009ApJ...703.1352K}  pointed out that radiation pressure depends on the radius of the ionized zone compared to a characteristic radius $\rch \propto L_*^2/S_* \propto L_*$ (for ionizing photon output $S_*\simeq L_*/(20\,\mbox{eV})$); \citet{2011ApJ...732..100D} had worked out the detailed structure; and \citet{2012ApJ...757..108Y} had explored the observational properties of similar regions enclosing stellar wind bubbles.  Using these developments MJ15 model ionized gas pressure as an enhancement to the radiation force and radiation-driven outflow rate, and demonstrated that the enhancement factor is only significant at low values of $\sigmac$ and for clouds massive enough to contain massive stars.  The net effect is an increase in the effective force (MJ15 eq.\ 33) that sets in for $\sigmac\ltsim 5-10\,$km\,s$^{-1}$ but shuts off for $\sigma\ltsim 1-2$\,km\,s$^{-1}$ due to the absence of ionizing stars.    (In fact, the boundaries of this region shift considerably depending on the most massive, most ionizing star.) 

\subsection{Stellar winds} 

Massive-star winds convert a portion of the radiation momentum into fluid momentum near the stellar surface.   Again there is a direct force (due to wind-cloud collision), which can be considered just a minor enhancement of the direct radiation pressure,  and an indirect force (due to the pressure of any trapped wind energy) which can conceivably become very large if wind energy is effectively trapped.   There is ample evidence from x-ray cooling radiation, as well as optical and infrared line ratios, that stellar wind energy is not trapped within giant H\,II regions \citep{2009ApJ...693.1696H,2012ApJ...757..108Y}.  While this does not guarantee that wind energy is escapes during the cluster-formation stage, \citet{2010ApJ...709..191M} present strong theoretical arguments that it does.  Accordingly, MJ15 simply account for wind force as a simple multiplicative factor, and fold this (along with photo-ionization) into the radiation force. 

\subsection{Core-collapse supernovae} 

Individual supernova explosions contain sufficient momentum to unbind several thousand solar masses of clump material, and with one supernova per $\sim 10^2\,M_\odot$ there can be many explosions in a massive cluster.  However, these occur {\em at least} several million years after star formation begins.  When growing a cluster of several thousand $M_\odot$ that samples the stellar mass function randomly (rather than making the massive stars first), waiting times are typically 10-20\,Myr -- much longer than the characteristic formation time of most clusters.  Therefore, supernovae can usually be ignored. 

\section{Parameter space, evolution, and comparison to observations} 

\begin{figure*}
\vskip -0.2cm
\centering
$\begin{array}{cc}
\includegraphics[angle=0,width=13.cm]{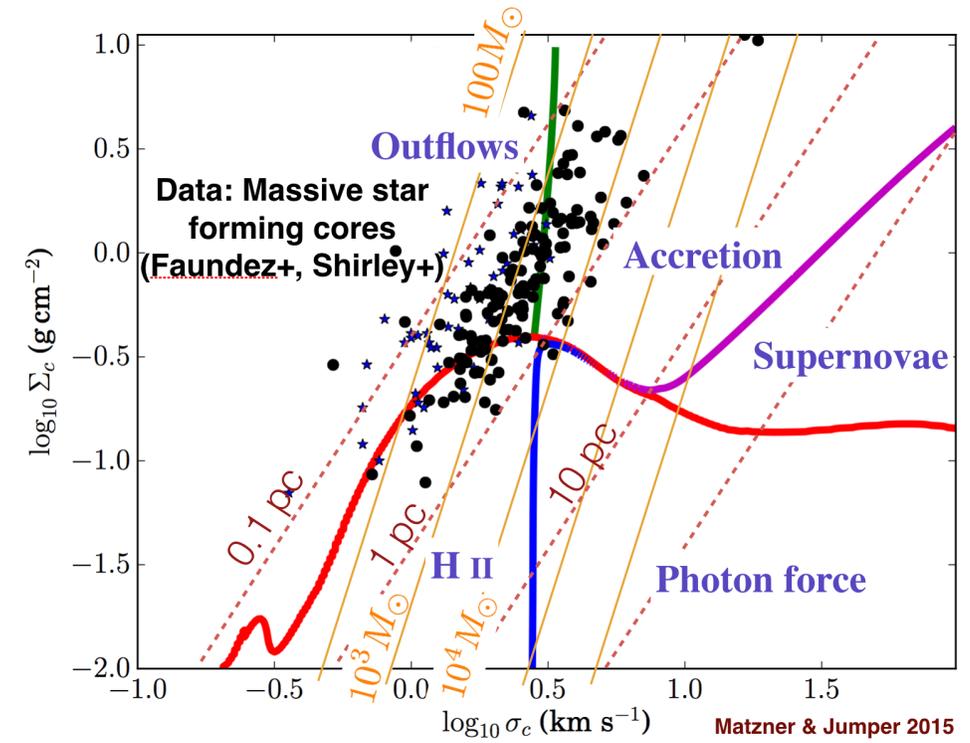} 
\end{array}$
\caption{
Parameter space of cluster-forming regions, adapted from MJ15.  Regimes of strong feedback from protostellar outflows, radiation pressure, and photo-ionization (H\,II) are shown; in the regime marked `accretion' none of these effects is significant.  Also shown are data from two surveys of massive star formation,  \citet{2003ApJS..149..375S} and \citet{2004A&A...426...97F}. }
\label{fig:ParamSpace}
\vspace{-0.5cm}
\end{figure*}

Combining all of the relevant effects -- accretion, star formation, protostellar outflows, radiation pressure, and (minimal) stellar winds -- it is possible to divide the parameter space of forming clusters into regimes based on $\Sigmac$ and $\sigmac$, as in Figure \ref{fig:ParamSpace}.   Protostellar outflows are important at low $\sigma$ ($\ltsim 3$\,km\,s$^{-1}$); these mainly erode and disrupt the clump, but cannot halt accretion or star formation \citep{2000ApJ...545..364M}.   Radiation pressure, and the effects that augment it, causes a rapid removal of residual gas if the column falls below $\sim 0.1$\,g\,cm$^{-2}$.   The intersection of these regimes also happens to be where photo-ionization boosts gas removal, raising the critical  $\Sigmac$ to about $0.3$\,g\,cm$^{-2}$.   

These regimes offer a convenient scenario for star cluster formation.  By driving down $\Sigmac$ at constant or increasing $\sigmac$, accretion forces cluster-forming clumps into a state where massive stars can remove the remaining gas.    In this case, one should expect to find regions down to the critical value of $\Sigmac$, but not below -- keeping in mind that the critical value is rather variable for stellar masses below $\sim 3000\,M_\odot$. 

However, the data in the  \citet{2003ApJS..149..375S} and \citet{2004A&A...426...97F} surveys tell a different story.   These regions appear to exist at nearly constant radius (about 0.3\,pc, with considerable scatter) regardless of mass.  As a result, low-mass regions skirt a low-$\sigmac$ region in which photoionization and outflows are important; but their massive counterparts extend into the high-$\sigmac$, high-$\Sigmac$ zone in which none of the feedback mechanisms identified here are strong.  

The clear conclusion for massive regions is that the mass reservoir, rather than feedback, brings cluster formation to an end.  Perhaps, as argued by  \citet{2016A&A...595A..27G} in the specific case of the W51 proto-clusters, feedback is more effective at removing the outlying reservoir material and starving the cluster of later accretion. 

For low-mass regions, it appears that photo-ionization and radiation pressure are significant.  Only a minority of these cross the boundary into the regime typically cleared of gas by these effects, and (as we have noted) that boundary is quite stochastic at low masses.   At face value this suggests that stellar feedback {\em is} important  at low masses (or slow cluster accretion rates).  However, the fact that the low-mass regions share a common radius with their high-mass cousins argues against this interpretation.   In any case, it is reassuring that there is no population clearly extending into the regime of rapid gas removal; this would indicate major errors in the estimation of photo-ionization feedback.     A future research project will more carefully review the sources that do cross the boundary of gas removal by photo-ionization, checking their individual luminosities and tracers of ionization, to more thoroughly compare them against theory. 

Finally, it should be noted that several low-mass regions are broadly consistent with MJ15's predictions where one can assess the effects of protostellar outflows.  MJ15 make detailed comparisons to the NGC\,1333 and Serpens South proto-clusters, finding that the velocity dispersion in the former can be understood as a combination of turbulent driving by accretion and by outflows, and that its gas and star content is consistent with expectations based on mass ejection, whereas Serpens South is in a young, filamentary dynamical state for which the theory does not apply.  In the region AFGL\,5142, \citet{2016ApJ...824...31L} observe a cluster-driven outflow that appears to be consistent with the mass loss expected in the theory.   

\section{Caveats and Discussion} 

The theory discussed here involves many simplifications, each of which merits its own caveats; but I will highlight only two.   First, there is the assumption that the reservoir material does not form stars before landing in the potential well of the cluster-forming clump.  There are certainly examples of filamentary structures within molecular clouds that appear to be stabilized by thermal pressure and transverse magnetic fields, and would survive the journey.  However, there are also structures that appear to be supported largely by turbulence, and some of these show star formation along their axes \citep{2011A&A...535A..76N}.  The Orion Nebula Cluster, for example, retains a kinematic fossil of its filamentary origin \citep{2011MNRAS.418.1796F}.   Second, the enterprise rests on the notion that the clump's dimensionless star formation rate $\SFRff$ is low enough, and constant enough, that  gas is trapped (and must virialize) while stars gradually form and feedback takes place.   If star formation occurred on one crossing time of the clump, this assumption and our predictions would break down.  

The analysis discussed here opens a number of  avenues for future investigation.  For our part, Peter Jumper and I are currently studying the details of dusty radiation transport (in idealized or realistic environments)  in order to specify the force enhancement factor more precisely; undergraduate Ziling Lu and I are re-examining the fates of stellar winds in cluster-forming regions.   Many numerical and observational follow-up projects are possible, from controlled experiments in the collapse of turbulent or filamentary regions, to careful comparisons of regions that display massive star formation against the MJ15 predictions for protostellar and ionized outflows.  

It is important to note that the MJ15 model is inspired by, and incorporates elements from, many previous theoretical works.  In addition to the models for protostellar and massive-star feedback mentioned above (which were already incorporated into the work of \citealt{2010ApJ...710L.142F}), MJ15's treatment of accretion-driven turbulence is influenced by \citet{2011ApJ...738..101G}, which itself was inspired by \citet{1999ApJ...527..285B}, \citet{2010A&A...520A..17K}, and \citet{2011MNRAS.411...65B}.   There are also similar analyses to be found in the analytical and theoretical work of  \citet{2012ApJ...746...75M}; \citet{2013prpl.conf1K105Z,2012ApJ...751...77Z}; and \citet{2016A&A...591A..30L,2016A&A...591A..31L}, among others.   We hope that the incorporation of both accretion and feedback, and the separation into physical regimes, will make MJ15's analysis useful for understanding observations of star cluster formation, for interpreting numerical simulations, and for motivating further, more detailed work in the field. 

\section{Acknowledgments}

My greatest thanks go to Quang Nguy$\tilde{\hat{\rm e}}$n L'o'ng, Fumitaka Nakamura, and the other organizers of the conference ``Star Formation in Different Environments'', at the very impressive ICISE facility in Quy Nhon, Vietnam, for their invitation, warm welcome, and effort.   The conference was a splendid example of the benefits of open borders and the free exchange of knowledge.   Of course I am also very grateful to Peter Jumper for the collaborative effort that led to MJ15 and our continuing research in the field of massive-star feedback.  I am supported by a Canadian NSERC Discovery Grant, and have recently benefited from the generosity, hospitality, and stimulating environment of the Monash Centre for Astrophysics in Melbourne, Australia. 

%




\end{document}